\documentclass[twocolumn,showpacs,twoside,10pt,aps,prl,superscriptaddress,longbibliography]{revtex4-2}
\usepackage{graphicx}
\usepackage{epsfig}
\usepackage{epsf}
\usepackage{amsmath,amsfonts,amssymb,natbib}
\usepackage{url,textcase,bm}
\usepackage{amsthm}
\usepackage{multirow}
\usepackage{mathrsfs}
\usepackage[colorlinks=true,citecolor=blue,urlcolor=blue]{hyperref}
\newcommand\rket[1]{|#1\rangle}
\newcommand\lket[1]{\langle #1|}
\begin{document}

\title{
Quantum Anomaly Detection with a Spin Processor in Diamond
}

\author
{
Zihua Chai$^{1,3\ast}$,
Ying Liu$^{1,3\ast}$,
Mengqi Wang$^{1,3}$,
Yuhang Guo$^{1,3}$,
Fazhan Shi$^{1,2,3}$,
Zhaokai Li$^{1,2,3\dag}$,
Ya Wang$^{1,2,3\dag}$,
Jiangfeng Du$^{1,2,3\dag}$
\\
\normalsize{$^{1}$ CAS Key Laboratory of Microscale Magnetic Resonance and School of Physical Sciences,}\\
\normalsize{University of Science and Technology of China, Hefei 230026, China.}\\
\normalsize{$^{2}$ Hefei National Laboratory, University of Science and Technology of China, Hefei 230088, China.}\\
\normalsize{$^{3}$ CAS Center for Excellence in Quantum Information and Quantum Physics, }\\
\normalsize{University of Science and Technology of China, Hefei 230026, China.}\\
\normalsize{$^{\ast}$ These authors contributed equally to this work.}\\
\normalsize{$^\dag$ E-mail: zkli@ustc.edu.cn, ywustc@ustc.edu.cn, djf@ustc.edu.cn}
}

\begin{abstract}
In the processing of quantum computation, analyzing and learning the pattern of the quantum data are essential for many tasks. Quantum machine learning algorithms can not only deal with the quantum states generated in the preceding quantum procedures, but also the quantum registers encoding classical problems. In this work, we experimentally demonstrate the anomaly detection of quantum states encoding audio samples with a three-qubit quantum processor consisting of solid-state spins in diamond. By training the quantum machine with a few normal samples, the quantum machine can detect the anomaly samples with a minimum error rate of 15.4\%. These results show the power of quantum anomaly detection in dealing with machine learning tasks and the potential to detect abnormal output of quantum devices.
\end{abstract}

\maketitle
Quantum computers have demonstrated their ability to deal with high-dimension data and corresponding linear algebra problems with quantum speed-up. Due to the natural essence as a linear system, a quantum register can efficiently interpret datasets stored as matrices and then perform matrix processing in the form of quantum algorithms \cite{harrow_quantum_2009, cai_experimental_2013, pan_experimental_2014}. This includes both the data encoding classical problem and the quantum states generated by a quantum device  \cite{aimeur2006machine}, such as the intermediate states in quantum many-body systems and quantum computational chemistry \cite{jiang_quantum_2018, mcardle_quantum_2020}. Since the complete classical description of a quantum register is usually hard to obtain and resource-consuming \cite{nielsen_2000}, the ability to investigate and process quantum data with a quantum computer is essential. In this process, quantum machine learning methods are promising tools to reveal the patterns of quantum data and use them in subsequent tasks, just as their counterparts running on classical computers \cite{paparo_quantum_2014, biamonte_quantum_2017,dunjko_machine_2018,schuld_quantum_2019, saggio_experimental_2021,huang_quantum_2022}. Many learning tasks, such as classification and feature extraction, have been demonstrated algorithmically or experimentally using quantum machine learning algorithms including principal component analysis \cite{lloyd_quantum_2014, li_resonant_2021}, support vector machines \cite{rebentrost_quantum_2014,li_experimental_2015}, and generative adversarial network models \cite{lloyd_quantum_2018,hu_quantum_2019,huang_experimental_2021}. With the ability to learn from quantum data, these methods integrate machine learning into quantum algorithms, constituting a great extension to the conventional algorithms on quantum computers.

Among the learning tasks in data processing, many appear in the form of anomaly detection (AD), i.e., the identification of the outliers appearing inconsistent with others in the datasets. 
It has broad application in both classical problems such as medical diagnosis and fraud detection in finance, and quantum problems including quantum state identification and phase diagram analysis \cite{hara_anomaly_2014, hara_quantum-state_2016, kottmann_unsupervised_2020, kottmann_variational_2021}.
In these problems, the normal cases are well sampled, while the anomalies are rare and under-sampled. 
This unbalance limits the performance of multi-class classification algorithms on these datasets, and gives rise to the research of anomaly detection algorithms \cite{chandola_anomaly_2009, pimentel_review_2014}.
Multiple classical algorithms are proposed from different perspectives, such as density estimation method by estimating the probability density function of the normal data \cite{breunig2000lof}, $k$-nearest neighbor method by calculating the distance of a sample to its $k$th nearest neighbor \cite{hautamaki_outlier_2004}, and one-class support vector machine by learning the boundary of the normal data \cite{choi_least_2009}. 
Recently, quantum algorithms for anomaly detection have been proposed \cite{liu_quantum_2018,liang_quantum_2019,herr_anomaly_2021,kottmann_variational_2021}, showing the potential to detect anomalies with resources growing logarithmically with respect to the number and dimension of training samples. 
The quantum algorithms can efficiently provide the anomaly score of a new test sample of interest, and then one can label the new sample as \textsc{normal} or \textsc{anomaly} according to the anomaly score.

In this work, we report an experimental demonstration of quantum anomaly detection (QAD) with a hybrid spin system in diamond at ambient conditions. By encoding classical data into the quantum processor, we apply the algorithm to an audio recognition problem as a full proof-of-principle demonstration of QAD. The quantum processor can calculate the inner products of multiple quantum samples in parallel and then use the results to identify whether a new audio sample is similar to the pattern of previously given samples.
After learning the distribution of the training samples in the feature space, our quantum processor can efficiently classify the test samples with a minimum error rate of 15.4\%. 

The task of the anomaly detection algorithm is to assign each new sample an anomaly score that can describe how far away it is from the pattern of normal samples. After this, one can set a threshold and classify a sample as \textsc{anomaly} if its anomaly score exceeds the threshold. Here the training set labeled as \textsc{normal} consists of $\mathrm{M}$ training samples $\vec{z}_{i}$ $\{i=1,2,...,\mathrm{M}\}$ and the new sample to be inspected is denoted by $\vec{z}_{\rm test}$. Each sample is represented by a vector in $\mathrm{d}$-dimension feature space and all the training samples are pre-centralized, i.e., $\sum_{i=1}^{\mathrm{M}} \vec{z}_{i}=\mathrm{0}$. 

A simple way to define the anomaly score is to compute the Euclidean distance between the test sample and the centroid of the training data \cite{hodge_survey_2004},
\begin{equation}
g\left(\vec{z}_{\rm test}\right)=\left|\vec{z}_{\rm test}\right|^{2}.
\label{eq:Euclidean_distance}
\end{equation}
A large Euclidean distance from the centroid predicts that the new sample is likely to be an \textsc{anomaly} case. The definition of Euclidean distance grants the same weight to the deviations in different dimensions, which makes it only work well when the samples have a nearly isotropic distribution in the feature space. On the other hand, if the training data are widely distributed in one direction while clustered in another, the Euclidean distance cannot recognize this anisotropy property. This leads to the possibility that the \textsc{normal} and \textsc{anomaly} samples may have similar Euclidean distances and thus be misclassified.

To overcome this, one needs to analyze the distribution of the training data in the feature space and quantify how well the inspected sample fits it. 
This can be done by using the anomaly score defined by proximity measure $f\left(\vec{z}_{\rm test}\right)$ \cite{liu_quantum_2018}, which compares the Euclidean distance $g\left(\vec{z}_{\rm test}\right)$ and the variance of the training data along the direction of $\vec{z}_{\rm test}$:
\begin{equation}
f\left(\vec{z}_{\rm test}\right)=\left|\vec{z}_{\rm test}\right|^{2}-\hat{z}_{\rm test}^{T} C \hat{z}_{\rm test}.\label{eq:proximity_measure_classical}
\end{equation}
Here $\hat{z}_{\rm test}=\vec{z}_{\rm test}/\left|\vec{z}_{\rm test}\right|$ and $C=\frac{1}{\mathrm{M}-1} \sum_{i=1}^{\mathrm{M}} \vec{z}_{i} \vec{z}_{i}^{T}$ is the covariance matrix which represents the distribution of all the training samples. 
The proximity measure can better reveal the pattern of training data, since it can provide a boundary line of \textsc{normal} data that fits the distribution of the training data better \cite{SM}. To obtain the proximity measure for a test sample, classical computers require $\mathrm{O(M*d)}$ resources to store and process the dataset, and costs $\mathrm{O(M*d)}$ computing resources to obtain the covariance matrix and do the following calculations.

\begin{figure}
\includegraphics[width=1\columnwidth]{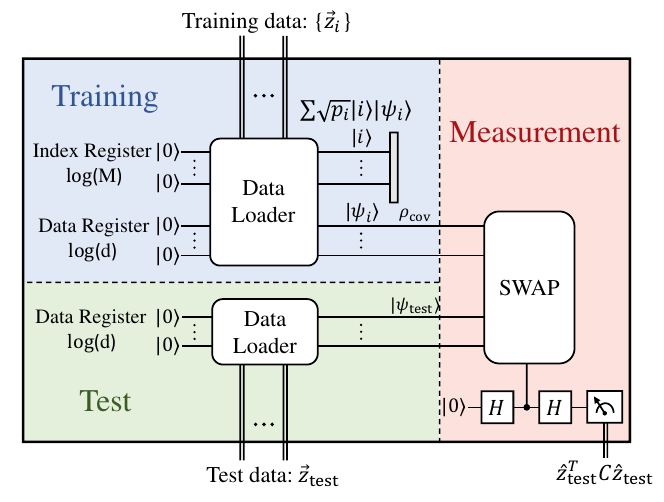}
\caption{
Schematic diagram of quantum anomaly detection algorithm. The training data and test data are loaded into the quantum registers and a \textsc{swap} test is used to estimate how close the test sample is from the pattern of the normal samples in the training set. The \textsc{swap} test could be replaced by other overlap-estimation methods, depending on the specific physical system used in the experiments.
}\label{fig1}
\end{figure}

A quantum computer can process the above samples and compute $\hat{z}_{\rm test}^{T} C \hat{z}_{\rm test}$ in Eq.\  \ref{eq:proximity_measure_classical} using resources logarithmic in the dimension ($\mathrm{d}$) and the number ($\mathrm{M}$) of the training samples \cite{liu_quantum_2018}. If the samples $\vec{z}_i$ are pure quantum states in $\mathrm{d}$-dimension Hilbert space, they can be directly loaded into the quantum register with standard \textsc{swap} gates. Here the quantum register is composed of a $\mathrm{log(d)}$-qubit data register and a $\mathrm{log(M)}$-qubit index register, as shown in Fig.~\ref{fig1}.
For classical data, a training-data oracle is employed to load the training samples into the quantum register. 
For each $\mathrm{d}$-dimension vector $\vec{z}_{i}$, the data loader returns a training state $\left|\psi_{i}\right\rangle=1 /\left|\vec{z}_{i}\right| \sum_{j=1}^{\mathrm{d}}\left(\vec{z}_{i}\right)_{j}|j\rangle$ and stores it into the data register. Here $|j\rangle$ represents the computational basis, and $\left(\vec{z}_{i}\right)_{j}$ represents the $j\mathrm{th}$ element of the $i\mathrm{th}$ sample. The fast loading of all the $\mathrm{M}$ vectors can be addressed by using quantum random access memory (QRAM) \cite{giovannetti_quantum_2008}. Starting with the superposition in the index register, i.e. $\frac{1}{\sqrt{\mathrm{M}}}\sum_{i=1}^{\mathrm{M}} |i\rangle$, the information of all training samples is stored as a superposition state
$\rket{\Psi}=\sum_{i=1}^{\mathrm{M}} \sqrt{p_{i}}\ |i\rangle\left|\psi_{i}\right\rangle$,
where $p_{i}=\frac{\left|\vec{z}_{i}\right|^{2}}{\sum_{i=1}^{\mathrm{M}}\left|\vec{z}_{i}\right|^{2}}$ represents the normalized module length $|\vec{z}_{i}|^2$ of the training data. At this stage, the reduced density matrix of the quantum state in the data register is 
\begin{equation}
\rho_\mathrm{cov}=\sum_{i=1}^{\mathrm{M}} p_{i}\left|\psi_{i}\right\rangle\left\langle\psi_{i}\right|,
\label{eq:reduce_mat}
\end{equation}
which has a similar form with the covariance matrix $C$ in Eq.\ \ref{eq:proximity_measure_classical}, i.e., $\rho_\mathrm{cov} = C/\xi$, where  $\xi=\frac{\sum_{i=1}^{\mathrm{M}}\left|\vec{z}_{i}\right|^{2}}{\mathrm{M}-1}$ is the trace of $C$. 

Following the same method, the test sample $\vec{z}_{\rm test}$ is loaded into the quantum register in the form of test state $\left|\psi_{\rm test}\right\rangle= \sum_{j=1}^{\mathrm{d}}\left(\hat{z}_{\rm test}\right)_{j}|j\rangle$. Then the proximity measure in Eq.\  \ref{eq:proximity_measure_classical} reduces to 
\begin{equation}
\begin{aligned}
f\left(\vec{z}_{\rm test}\right) =  \left|\vec{z}_{\rm test}\right|^{2}- \xi \  \lket{\psi_{\rm test}} \rho_{\mathrm{cov}} \rket{\psi_{\rm test}}.
\label{eq:quantum_proximity}
\end{aligned}
\end{equation}
By measuring the overlap between $\rho_{\mathrm{cov}}$ and $\rket{\psi_{\rm test}}$, the inner products of the test state and all the training states are calculated simultaneously, leading to a fast estimation of the proximity measure. In the experiments, one can store $\rho_{\mathrm{cov}}$ and $\rket{\psi_{\rm test}}$ into different quantum registers and then adopt overlap estimation methods such as the \textsc{swap} test \cite{buhrman_quantum_2001}.
In this case, the probability of the ancillary qubit at $\rket{0}$ returns $P_{\rket{0}}=(1+\lket{\psi_{\rm test}} \rho_{\rm cov} \rket{\psi_{\rm test}})/2$, which can be utilized to estimate the overlap needed and then obtain the anomaly score in Eq.\  \ref{eq:quantum_proximity}, as shown in Fig.~\ref{fig1}.

\begin{figure}
\includegraphics[width=1\columnwidth]{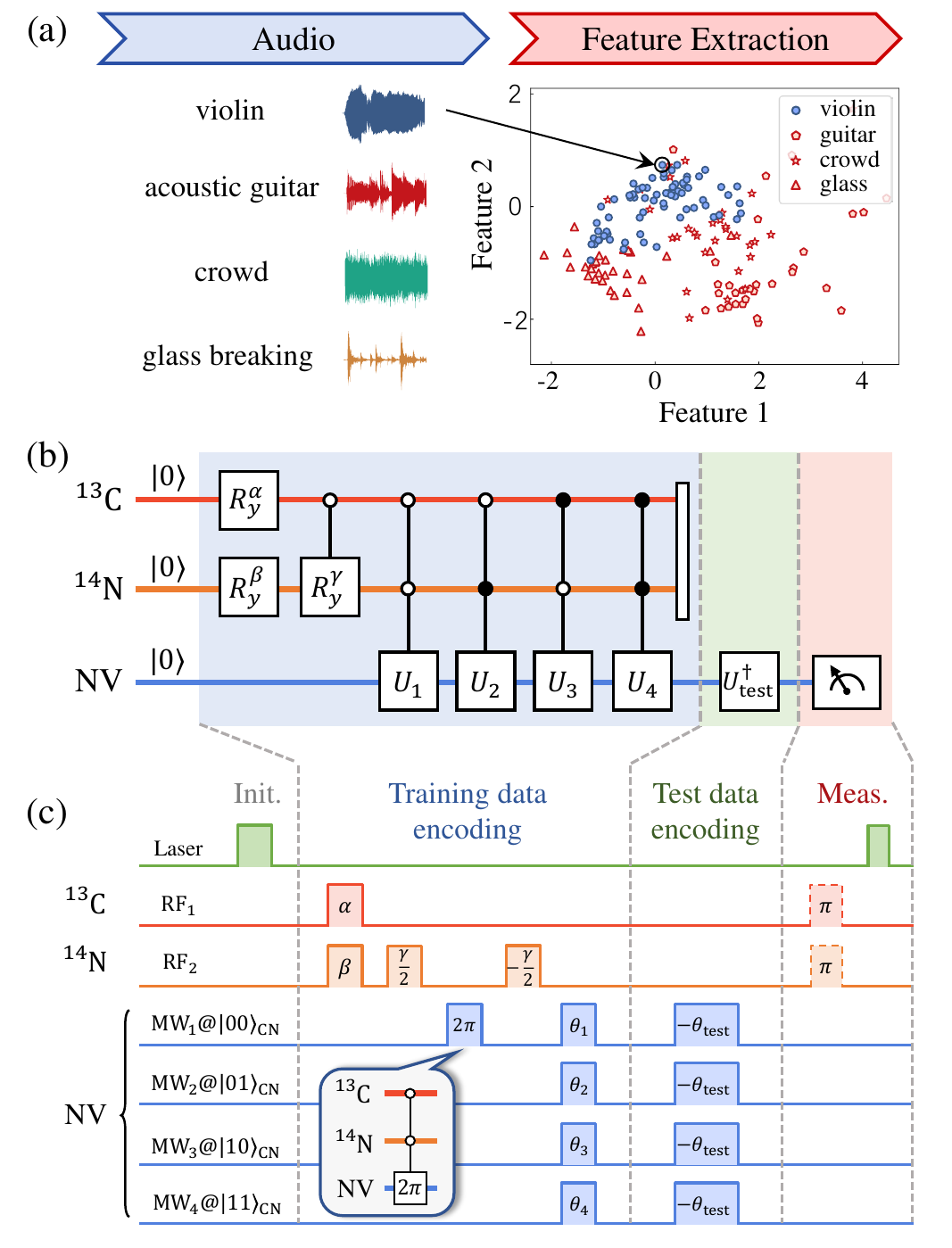}
\caption{
(a) The pre-processing of the audio samples. Waveforms of different types of audio samples are processed into feature vectors in the 2-dimension feature space. 
(b) The schematic diagram of the quantum anomaly detection in this work. $^{13}$C and $^{14}$N nuclear spins are utilized as the index register, while the electron spin (NV) is used to encode each sample conditional on the state of the index register. After encoding the test sample with $U_{\rm test}^{\dagger}$, the proximity measure is estimated by measuring the probability of the electron being on state $\rket{0_\mathrm{e}}$.
(c) Pulse sequence used in the experiments. The non-local gate $\mathrm{C}^{\rm C} \mathrm{ROT}^{\rm N}(\gamma)$ in (b) is realized by the conditional phase gate on the electron spin and two local operations, where the second radio-frequency pulse is $\pi$-phase shifted relative to the first one. The data-loading operations $\rket{i}\lket{i}\otimes U_i$ are conducted in parallel by applying the microwave pulses with an arbitrary wave generator.
}\label{fig2}
\end{figure}

As an application, we apply the QAD scheme to an audio recognition problem. This demonstration aims to identify whether a piece of audio segment belongs to a certain type of sound, e.g., the sound of the violin in our case. The acoustic samples we used come from the dataset for acoustic event recognition in Ref.\ \cite{takahashi_deep_2016}. Here we choose four types of audio samples: \{\textit{violin}, \textit{acoustic guitar}, \textit{crowd} and \textit{glass breaking}\} as the dataset, with \{70, 30, 30 and 30\} samples of each type, respectively. The original waveforms are firstly analyzed by the method of Mel-frequency cepstral coefficients (MFCCs) \cite{sahidullah_design_2012,muller_information_2007}, and then processed by the feature extraction method to reduce the dimension of each sample while keeping most of the information \cite{roma2019adaptive, dupont_nonlinear_2013}.
Details of the pre-processing are shown in Supplemental Material \cite{SM}. After pre-processing, each audio segment is represented by a 2-dimension feature vector and is ready to be analyzed by quantum processor. In our demonstration, the training set labeled as \textsc{normal} is composed of the following elements 
\begin{equation}
\begin{aligned}
& \vec{z}_{1}=\left(-0.789,0.130\right), \vec{z}_{2}=\left(0.751,-0.023\right),\\
& \vec{z}_{3}=\left(0.617,0.531\right), \vec{z}_{4}=\left(-0.579,-0.639\right),
\end{aligned}
\end{equation}
and has very similar covariance matrix with the dataset consisting of all the violin samples.

The quantum processor used in this experiment is a nitrogen-vacancy defect (NV) center electron spin (S=1) associated with the intrinsic nitrogen nuclear spin (N, S=1) and a nearby carbon nuclear spin (C, S=1/2). The subspace of $\{m_\mathrm{e}=0,+1\}\otimes\{m_\mathrm{C}=+1/2,-1/2\}\otimes\{m_\mathrm{N}=+1,0\}$ forms a three-qubit system, which is labeled as $\{\rket{0},\rket{1}\}\otimes\{\rket{0},\rket{1}\}\otimes\{\rket{0},\rket{1}\}$. 
In the experiment, the electron spin is chosen as the data register to store the training states $\vec{z}_{i}$, due to its fast, versatile, and high-fidelity control and readout \cite{neumann_single-shot_2010,robledo_high-fidelity_2011,dolde_room-temperature_2013,dolde_high-fidelity_2014,rong_experimental_2015,zhang_high-fidelity_2021}. The two nuclear spins serve as the index register to store the label $|i\rangle$ of each sample and the normalized module length of training data, utilizing the advantage of long coherence time \cite{maurer_room-temperature_2012,yang_high-fidelity_2016,bradley_ten-qubit_2019}.  

Starting from the thermal state, the three-qubit quantum system is initialized to state $\rket{000}$ by a short laser pulse for dynamical nuclear polarization \cite{jacques_dynamic_2009, smeltzer_13_2011}. To load the data of audio samples into the quantum processor, we begin with preparing the superposition state in the index register, i.e., $\sum_{i=1}^{4} \sqrt{p_{i}}\ |i\rangle$, which is weighted by the normalized module length $p_i$ of each training sample. This is realized by a parameterized quantum circuit composed of single-qubit rotation $R_y^{\rm C}(\alpha)$, $R_y^{\rm N}(\beta)$, and a non-local gate $\mathrm{C}^{\rm C} \mathrm{ROT}^{\rm N}(\gamma)=\rket{0}_{\mathrm{C}}\lket{0}\otimes R_y(\gamma)+\rket{1}_{\mathrm{C}}\lket{1}\otimes{\rm{I}_2}$, which denotes a single-qubit rotation $R_y^{\rm N}(\gamma)$ conditioned on the carbon nuclear spin being at state $\rket{0}_{\rm C}$. Here $R_y(\theta) = e^{-i{\sigma_y} \theta/2}$. 
The non-local gate between the nuclear spins is composed of the conditional $\pi$-phase (CPhase) gate on the electron spin and two local operations on the nitrogen nuclear spin \cite{waldherr_quantum_2014}. By addressing the hyperfine splitting, a frequency-selective microwave pulse is applied to the electron spin. The microwave pulse rotates the electron spin with an angle of $2\pi$ conditional on $\rket{0}_{\mathrm{C}}\rket{0}_{\mathrm{N}}$ ($\mathrm{MW}_1@\rket{00}_{\rm CN}$), leading to the CPhase gate, i.e. the conditional $\pi$ phase shift $\rket{0}_{\mathrm{e}}\rket{0}_{\mathrm{C}}\rket{0}_{\mathrm{N}} \rightarrow e^{i\pi}\rket{0}_{\mathrm{e}}\rket{0}_{\mathrm{C}}\rket{0}_{\mathrm{N}}$.

After having $\sum_{i=1}^{4} \sqrt{p_{i}}\ |i\rangle$ in the index register, the next step is to load the information of each $\vec{z}_{i}$ into the data register conditioned on the state $\rket{i}$ in the index register, i.e.\ to perform  $\sum_{i=1}^{4} \rket{i}\lket{i}\otimes U_i$ (Fig.~\ref{fig2}(b)), where $U_i\rket{0}=\rket{\psi_{i}}$, representing the encoding operation. In the experiment, the encoding of $\vec{z}_{i}$ can be achieved by applying the selective microwave pulse at frequency ${\rm MW}_i$, which rotates the electron spin with the angle $\theta_i=2 \mathrm{arctan}({\left(\vec{z}_{i}\right)_2}/{\left(\vec{z}_{i}\right)_1})$ (Fig.~\ref{fig2}(c)). With an arbitrary wave generator compiling all the encoding microwave pulses into a single pulse, we prepare different $\rket{\psi_{i}}$ in parallel and obtain the superposition of the training states, i.e. $\rket{\Psi}=\sum_{i=1}^{\mathrm{4}} \sqrt{p_{i}}|i\rangle\left|\psi_{i}\right\rangle$. 
After discarding the index register, the density matrix of the data register (electron spin) is  $\rho_\mathrm{e}=\rho_\mathrm{cov}=C/\xi$. 
When solving a problem with larger amounts of data, the quantum circuit here for data loading and analysis can be generalized. Furthermore, QRAM can be used to reduce the data loading complexity.

In the readout process, the proximity measure of a new test sample $\vec{z}_{\rm test}$ is estimated by measuring the overlap between the state in the data register $\rho_\mathrm{e}$ and the test state $\left|\psi_{\rm test}\right\rangle$, i.e. $\lket{\psi_{\rm test}} \rho_{\rm cov} \rket{\psi_{\rm test}}$ in Eq.\ \ref{eq:quantum_proximity}. Although the method of \textsc{swap} test can estimate the proximity measure of test samples, it is at the cost of more ancillary qubits. In the experiment, we use another efficient approach to introduce the information of the test sample $\vec{z}_{\rm test}$, utilizing the inverse operation of the data loading, denoted by $U_{\rm test}^{\dagger}$. The overlap $\lket{\psi_{\rm test}} \rho_{\mathrm{e}} \rket{\psi_{\rm test}}$ can be estimated by a single measurement of the population of the electron spin at $\rket{0}_{\mathrm{e}}$ after applying $U_{\rm test}^{\dagger}$, since $\lket{\psi_{\rm test}} \rho_{\mathrm{e}} \rket{\psi_{\rm test}}=\lket{0} U_{\rm test}^{\dagger} \rho_{\mathrm{e}} U_{\rm test} \rket{0}$. The population is read out by counting the number of photons emitted from the NV center after optical pumping with a short laser pulse (see Supplemental Material \cite{SM}). Combined with the module lengths of the training vectors and the test vector, the proximity measure $f\left(\vec{z}_{\rm test}\right)$ is obtained. 

\begin{figure}
\includegraphics[width=1\columnwidth]{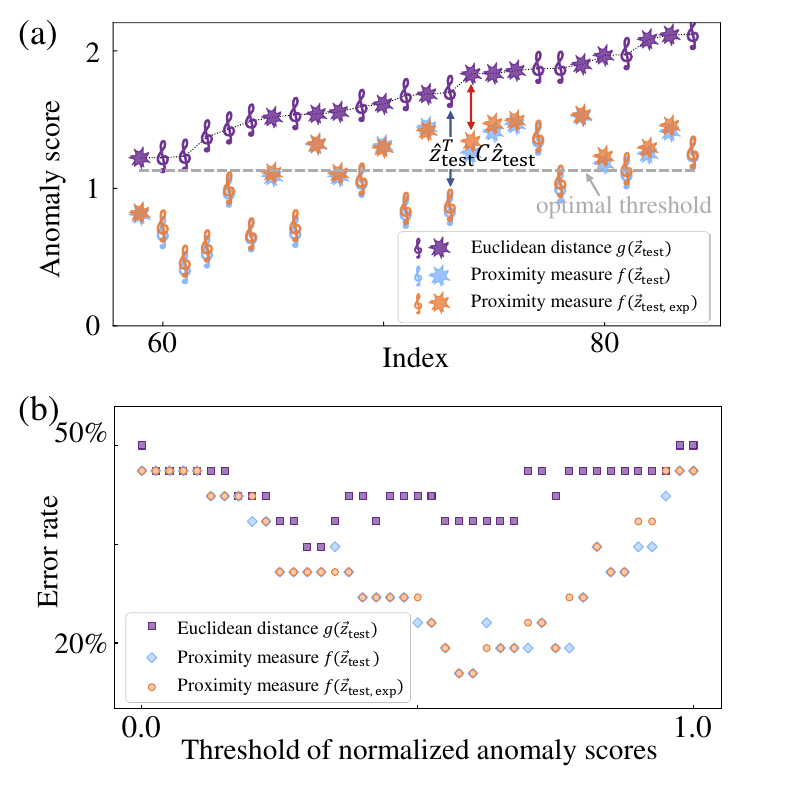}
\caption{Anomaly detection by the methods of Euclidean distance $g\left(\vec{z}_{\rm test}\right)$ and proximity measure $f\left(\vec{z}_{\rm test}\right)$. Here, the music notes and stars represent normal (violin) samples and anomaly (non-violin), respectively.  (a) Anomaly scores provided by different methods. The samples are sorted in ascending order of $g\left(\vec{z}_{\rm test}\right)$. Error bars are not shown since they are much smaller than the size of the markers. (b) The error rate of different methods.  The threshold is scanned within the range of anomaly scores in (a), after the range is normalized to $[0, 1]$.
}\label{fig3}
\end{figure}

In the experiments, we perform proximity measurement on 111 test samples of different types of audio. The samples are re-ordered in the ascending order of their Euclidean distance $g\left(\vec{z}_{\rm test}\right)$ from the centroid of the data. When the test sample is far away (or very close) to the centroid, it is trivial to classify it as \textsc{anomaly} (or \textsc{normal}). We will focus on the overlapping area where the two kinds of samples are mixed and hard to distinguish, i.e., the area shown in Fig.~\ref{fig3}(a). Here the purple and blue symbols represent the Euclidean distance and the proximity measure calculated with Eq.\  \ref{eq:Euclidean_distance} and Eq.\  \ref{eq:proximity_measure_classical}, respectively. The orange symbols represent experimentally measured proximity $f\left(\vec{z}_{\rm test, exp}\right)$, which accord with the theoretical expectation well. The experimental results show that the measured proximity tends to have a higher value for the anomaly samples (non-violin) and lower for normal samples (violin), even when these two kinds of samples sometimes have similar Euclidean distances from the centroid. 

After having the anomaly scores of these samples, one can set a threshold and assign the label \textsc{anomaly} to the samples with scores greater than the threshold. Then the performance of each anomaly detection method can be evaluated using the error rate, which denotes the proportion of the mislabeled samples in the test set. Fig.~\ref{fig3}(b) shows how the error rate varies with the threshold. The horizontal axis is re-normalized to have 0 representing the lowest anomaly score in Fig.~\ref{fig3}(a) and 1 representing the highest one. It shows that the proximity measure offers a much lower error rate than the method of Euclidean distance in most of the area. The minimum error rate of our experimental quantum anomaly detection is 15.4\%, which is 55.6\% lower than the best performance that the Euclidean distance method can achieve (34.6\%). These experimental results show that our quantum processor can efficiently learn the distribution of the training samples and classify different test samples accordingly.

\begin{figure}
\includegraphics[width=1\columnwidth]{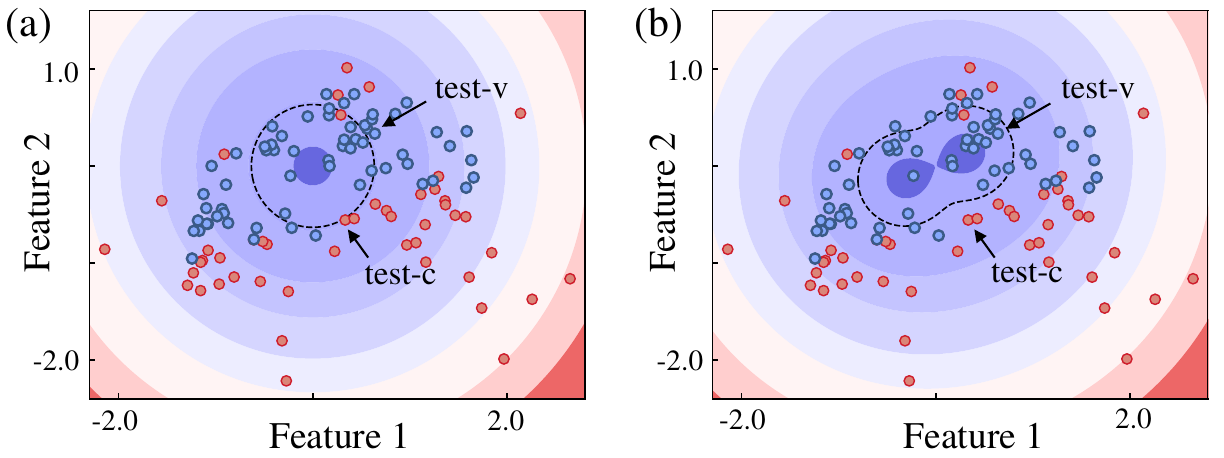}
\caption{The spatial distribution of (a) Euclidean distance and (b) proximity measure in the feature space. Areas in red have a higher probability to be \textsc{anomaly}, when areas in blue are more likely to be \textsc{normal}. Proximity measure gives a peanut-shaped boundary line which fits better with the distribution of the \textsc{normal} samples.
}\label{fig4}
\end{figure}

To better understand how the spatial distribution of the training samples is learned, we can read out the proximity measure in the feature space through the covariance matrix stored in the electron spin. 
The experimental reconstructed density matrix reads as
\begin{equation}
\begin{aligned}
\rho_{\mathrm{exp}}=\left(\begin{array}{cc}
0.6996 & 0.2151 \\
0.2151 & 0.3004 
\end{array}\right), 
\end{aligned}
\end{equation}
which matches the theoretical expectation $\rho_{\mathrm{cov}}$ with fidelity $F(\rho_\mathrm{exp}, \rho_\mathrm{cov}) = \mathrm{\textbf{tr}} \sqrt{\sqrt{\rho_\mathrm{exp}} \rho_\mathrm{cov} \sqrt{\rho_\mathrm{exp}}}=99\%$. For any point in the feature space $\vec{v}= (\rm{feature 1, feature 2})$, its proximity measure is estimated as $f\left(\vec{v}\right)=\left|\vec{v}\right|^{2}- \xi \  \lket{\psi_{v}} \rho_{\mathrm{exp}} \rket{\psi_{v}}$. Then the value of proximity measure is shown in Fig.~\ref{fig4}(b) represented by different colors in the feature space, while the value of Euclidean distance is shown in Fig.~\ref{fig4}(a).
In the figure of proximity measure, the peanut-shaped boundary line shows that it can better learn the distribution of the normal data (blue)  than using Euclidean distance. In the direction where the normal samples are widely distributed, the proximity measure $f\left(\vec{z}_{\rm test}\right)$ rises slowly with the distance from the centroid. For example, while test-v is further from the centroid than test-c (Fig.~\ref{fig4}(a)), it has a lower proximity measure than test-c (Fig.~\ref{fig4}(b)). This indicates that it is less likely to be an \textsc{anomaly} sample, which is in accordance with its label in the dataset. In this case, the \textsc{anomaly} samples are distributed in all directions around the normal samples, which is beyond the applicability of one-class support vector machine. 

In conclusion, we demonstrated a quantum algorithm for anomaly detection on a real quantum processor, i.e., the three-qubit hybrid spin system in diamond. Although the demonstration here is implemented in a classical dataset, the method itself can be adopted to identify quantum data, i.e. quantum states. In this work, we utilized the spin processor to demonstrate a full quantum anomaly detection process.
The quantum machine was trained with audio samples which are labeled as \textsc{normal}, and then estimated the anomaly scores of the new samples. The experimental results show that this quantum method efficiently analyzes the distribution of the training set in the feature space, which leads to the lower error rate in detecting the anomalies.  
Given a larger quantum processor, our work can be naturally extended to deal with more complicated classical problems efficiently, such as credit card fraud analysis with much more samples of higher dimensions.
Furthermore, a quantum processor can efficiently receive and process samples transported by quantum internet (e.g., by receiving photons containing quantum states). Thus this method can work as a subroutine to detect anomalies in quantum devices, avoiding the resource-consuming process of reading out high-dimension quantum states using quantum tomography or similar methods.

\section*{Materials and Methods}

\subsection{Experimental Setup}

The experiments were implemented using an NV center with a proximal $^{13}$C nuclear spin and an intrinsic $^{14}$N nuclear spin in a [100]-oriented diamond. In the rotating frame, the effective Hamiltonian of this spin system reads as
\begin{equation}
\begin{aligned}
\mathcal{H}_{\text{NV,eff}}=\rket{1}_{\mathrm{e}}\lket{1}\otimes(A_\parallel^{\mathrm{C}} \sigma_{z}^{\mathrm{C}}/2 + A_\parallel^{\mathrm{N}} \sigma_{z}^{\mathrm{N}}/2),
\end{aligned}
\end{equation}
where $\sigma_{x,y,z}$ are Pauli operators. $A_\parallel^{\mathrm{C}} \approx 12.8\ \mathrm{MHz} $ and $A_\parallel^{\mathrm{N}} \approx -2.16\ \mathrm{MHz} $ are the hyperfine coupling strengths between the electron spin and the two nuclear spins. The dephasing times of the spins are measured as $T_{2,\rm e}^* \approx $ 5.4 $\mu$s, $T_{2,\rm C}^* \approx $ 2.0 ms  and $T_{2,\rm N}^* \approx $ 5.0 ms. An external magnetic field of around 510 Gauss was applied by a permanent magnet along the symmetry axis of the NV center, so that the three-qubit system can be efficiently polarized to \{$m_{\mathrm{e}}=0, m_{\mathrm{C}}=+1/2, m_{\mathrm{N}}=+1\}$ after the polarization transfer in the excited state during laser pumping. 

We carried out the experiment on a home-built confocal microscope at ambient conditions. The optical pumping and readout of the electron spin are realized by a continuous-wave laser at 532 nm which is gated with two acoustic-optic modulators (AOM). The laser beam was focused by an oil objective, while the fluorescence signal was collected by the same objective. A solid immersion lens was fabricated on the NV center to improve the photon collection efficiency. An active temperature control to within 5 mK was used to increase the magnetic field stability. The microwave and radio-frequency signals used to control the electron spin were generated by an arbitrary waveform generator (AWG, crs1w000b, CIQTEK) in combination with a microwave generator through the I/Q modulation. The same AWG also generated the radio-frequency signal to control the nuclear spins.

\begin{figure}
\centering
\includegraphics[width=1\columnwidth]{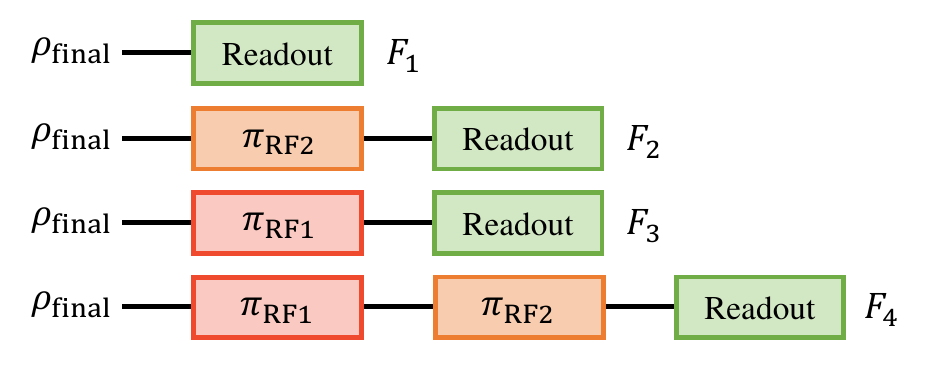}
	\caption{The readout of the possibility of electron spin at $\rket{0}_{\mathrm{e}}$. Here $\rho_{\mathrm{final}}$ is the three-qubit state to be measured. $\pi_{\mathrm{RF1}}$ and $\pi_{\mathrm{RF2}}$ represent single-qubit $\pi$ pulse on carbon spin and nitrogen spin, respectively. $F_i\ (i=1,2,3,4)$ is the number of photons detected in corresponding experiment.}\label{figs3}
\end{figure}

\subsection{Experimental Readout Method} 

To read out $\lket{\psi_{\rm test}} \rho_{\mathrm{e}} \rket{\psi_{\rm test}}$ for each test sample, we need to measure the possibility of electron spin at $\rket{0}_{\mathrm{e}}$, denoted as $P_0$. In the three-qubit system, $P_0$ is the sum of the populations at corresponding sublevels, i.e., $P_0=\Sigma_{m=0}^{3}\ p_m$ (accordingly $P_1=\Sigma_{m=4}^{7}\ p_m$). Here, $m$ labels the binary representation of the eight energy levels (e.g.\@ $\rket{0}=\rket{0_{\mathrm{e}} 0_{\mathrm{C}} 0_{\mathrm{N}}}$), and $p_m$ is the corresponding population. At the magnetic field of 510 Gauss, these energy levels give rise to different photon luminescence rates, leading to different detected photon numbers $N_m$.
Thus, the number of photons detected in the experiment, denoted by $F_1$, is related to the distribution of populations as $F_1=\Sigma_{m=0}^{7}\ p_m N_m$.
However, since different sets of distribution ${p_m}$ can produce the same $F_1$, one can not determine $P_0$ through a single readout of $F_1$. To overcome this, a series of experiments with different pulse sequences are required.

We apply the $\pi$ pulses shown in Fig.~\ref{figs3} to switch the population of different sublevels, and then count the number of photons in the new distribution. The numbers of photons detected in the four experiments give rise to the following equations

\begin{equation*}\label{p0_eq}
	\begin{aligned}
		\left(\begin{array}{l}
			F_1 \\
			F_2 \\
			F_3 \\
			F_4
		\end{array}\right)=\left(\begin{array}{cccccccc}
			p_0 & p_1 & p_2 & p_3 & p_4 & p_5 & p_6 & p_7 \\
			p_1 & p_0 & p_3 & p_2 & p_5 & p_4 & p_7 & p_6 \\
			p_2 & p_3 & p_0 & p_1 & p_6 & p_7 & p_4 & p_5 \\
			p_3 & p_2 & p_1 & p_0 & p_7 & p_6 & p_5 & p_4 
		\end{array}\right)\left(\begin{array}{l}
			N_0 \\
			N_1 \\
			N_2 \\
			N_3 \\
			N_4 \\
			N_5 \\
			N_6 \\
			N_7 
		\end{array}\right).
	\end{aligned}
\end{equation*}
The average of the photon numbers
\begin{equation*}
	\begin{aligned}
		\frac{F_1+F_2+F_3+F_4}{4} = & P_0\times\frac{N_0+N_1+N_2+N_3}{4}\\
		& +P_1\times\frac{N_4+N_5+N_6+N_7}{4},
	\end{aligned}
	\label{p0_exp}
\end{equation*}
effectively averages the photon luminescence rates of different sublevels, and thus leads to the readout of $P_0$. Then the proximity measure of the test sample $\vec{z}_{\rm test}$ can be obtained after having $P_0$.

\

\textit{Key words.---} quantum algorithm, quantum computing, machine learning, anomaly detection, nitrogen-vacancy center

\textit{Conflict of interest.---} The authors declare no conflict of interest.

\textit{Data Availability. ---} 
All data needed to evaluate the conclusions in the paper are present in the paper.

\textit{Author contributions. ---} J.D., Z.L., and Y.W. supervised the experiments. Z.L. and Y.W. proposed the idea and designed the experiments. Z.C. and Y.L. performed the experiments. M.W. fabricated the structure. Z.L., Z.C., and Y.W. analyzed the data. Y.W., Z.L., and Z.C. wrote the manuscript. All authors discussed the results and commented on the manuscript.

\begin{acknowledgments}
\textit{Acknowledgements.---} The authors thank Nana Liu for helpful discussions. 
This work was supported by the National Natural Science Foundation of China (Grants No.\ 92265204, 92165108), the Chinese Academy of Sciences (Grant No.\ GJJSTD20200001), the Anhui Provincial Natural Science Foundation (2108085J04), the Innovation Program for Quantum Science and Technology (Grant No.\ 2021ZD0302200), the Anhui Initiative in Quantum Information Technologies (Grant No.\ AHY050000), the Fundamental Research Funds for the Central Universities, and the University of Science and Technology of China (USTC) Research Funds of the Double First-Class Initiative.

\end{acknowledgments}

\bibliography{references_QAD}

\clearpage
\newpage

\begin{center}
    \Large
    Supplemental Material for "Quantum Anomaly Detection with a Spin Processor in Diamond"
    \\[10pt]
\end{center}
	
	\renewcommand{\thefigure}{S\arabic{figure}}
	\setcounter{figure}{0}

\section{Anomaly Detection Methods}
\begin{figure}[h]
	\includegraphics[width=0.8\columnwidth]{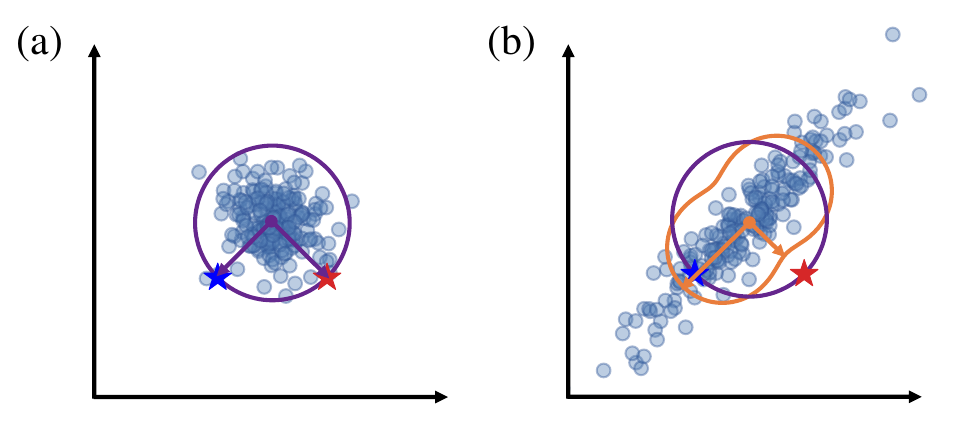}
	\caption{The boundary line given by different methods used as anomaly scores. Blue dots are the training samples labeled as \textsc{normal} in the feature space. The two stars are two test samples that have the same Euclidean distances with the centroid. The purple and orange lines represent the boundary lines given by certain values of Euclidean distance and proximity measure, respectively. (a) Nearly isotropic distribution in the feature space. 	(b) Anisotropic distribution in the feature space. Here the two test samples (stars) have the same Euclidean distances but rather different proximity measures. 
	}\label{figs0}
\end{figure}

Both Euclidean distance and proximity measure can be used as anomaly score for a test sample. If the distribution of the training samples is nearly isotropic in the feature space, the Euclidean distance between the test sample and the centroid can be used as the anomaly score (Fig.~\ref{figs0}(a)). The test sample is classified as an anomaly when the distance exceeds a threshold. However, if the training samples have different distributions in different directions, the Euclidean distance cannot recognize this anisotropy property (Fig.~\ref{figs0}(b)). To overcome this, the definition of proximity measure utilizes the spatial distribution of the training samples by introducing the covariance matrix. Thus when two test samples have similar Euclidean distances, they can have different proximity measures. In the direction where the normal samples are widely distributed, the proximity measure rises slowly with the distance from the centroid, leading to a lower probability of a test sample being classified as an anomaly, as shown in Fig.~\ref{figs0}(b).

\begin{figure}[h]
	\includegraphics[width=1\columnwidth]{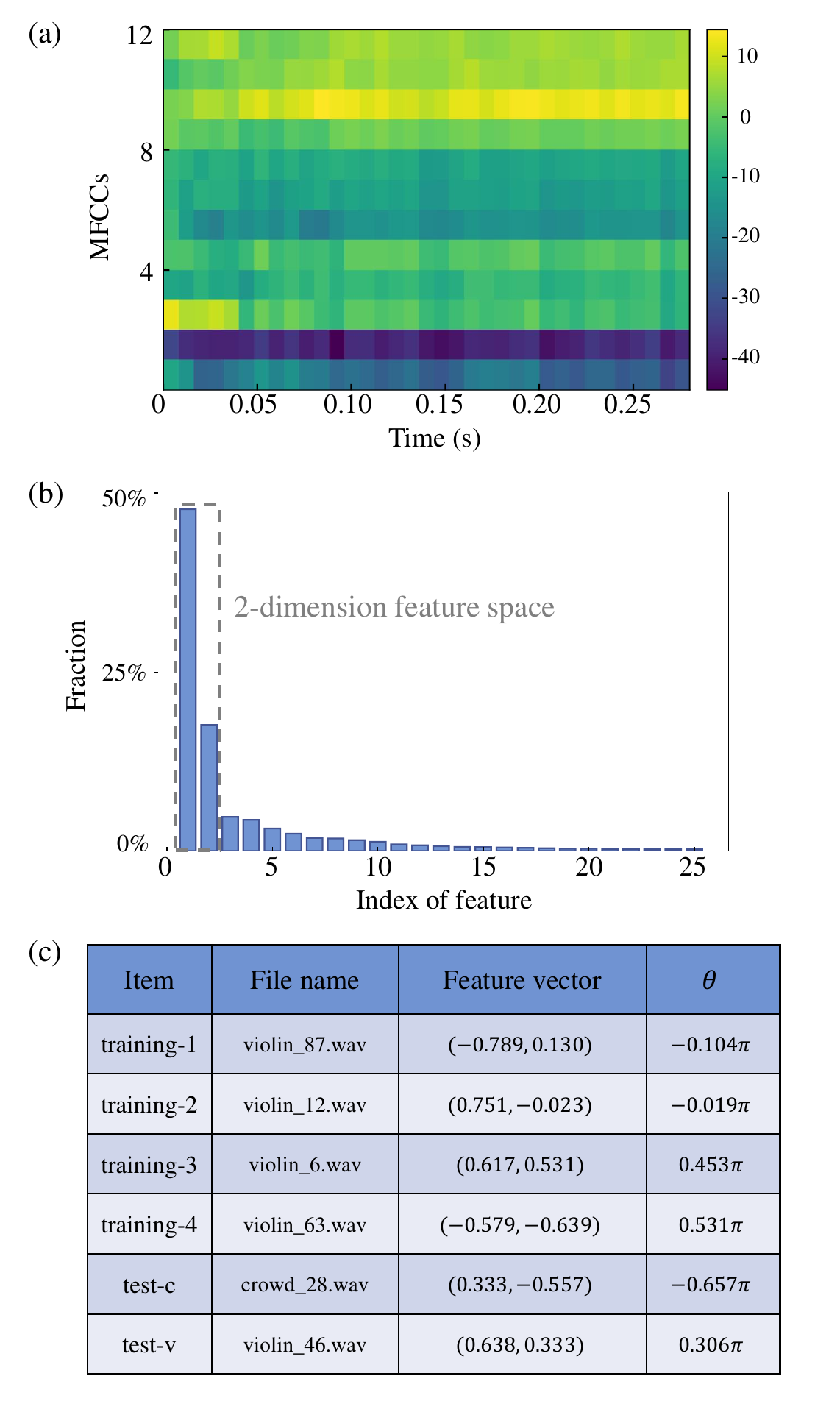}
	\caption{(a) Cepstral coefficients array extracted from violin-47.wav as an example. Each column represents the MFCCs extracted from one frame of the audio sample. (b) The proportion of variance explained by each feature in the principal component analysis. The first 25 principal components are shown. (c) The 2-dimension feature vectors of all the training samples and the two test samples mentioned in the main text. 
	}\label{figs1}
\end{figure}
\section{Dataset of the audio samples}
The original audio waveforms are collected from the dataset mentioned in the main text. After excluding the silence segments, we extract a 0.28-second waveform segment from each audio. Firstly we analyze the original waveform using the method of Mel-frequency cepstral coefficients, which can extract the properties of the short-term power spectrum and capture the timbral characteristics. The calculation of Mel-frequency cepstral coefficients consists of four steps: (i) dividing the audio segment into frames, (ii) taking the Fourier transform, (iii) mapping the powers of the spectrum to the Mel scale and taking the logs, and (iv) taking the discrete cosine transform. The segment of each audio sample is divided into 35 frames, and from each frame 12 cepstral coefficients are extracted. Here the length of each frame is 16 milliseconds, having 8 milliseconds overlap with adjacent frames. After this, each audio sample is processed into a 12 by 35 coefficients array (Fig.~\ref{figs1}(a)).

\newpage

To further project the feature vector into a 2-dimension space, we flatten each array to a 420-dimension vector and apply principal component analysis to extract the main features while keeping most of the information. Defining the eigenvalues of the covariance matrix of the dataset as $\lambda_{k}$, the proportion of variance explained by the two principal components reaches $\sum_{k=1}^{2}\lambda_{k}/{\sum_{k=1}^{420}\lambda_{k}} = 65\%$ (Fig.~\ref{figs1}(b)). Thus, the subspace spanned by the first and the second principal components reserves most of the information after feature extraction. We use these two principal components as the basis of the feature space, and the corresponding coordinate is (feature 1, feature 2), after being divided by 100 to rescale. The distribution of the audio samples in the feature space is shown in Fig.~2(a) in the main text. The corresponding vectors of all training samples and the two test samples mentioned in the main text are shown in Fig.~\ref{figs1}(c).

\begin{figure}[h]
	\includegraphics[width=1\columnwidth]{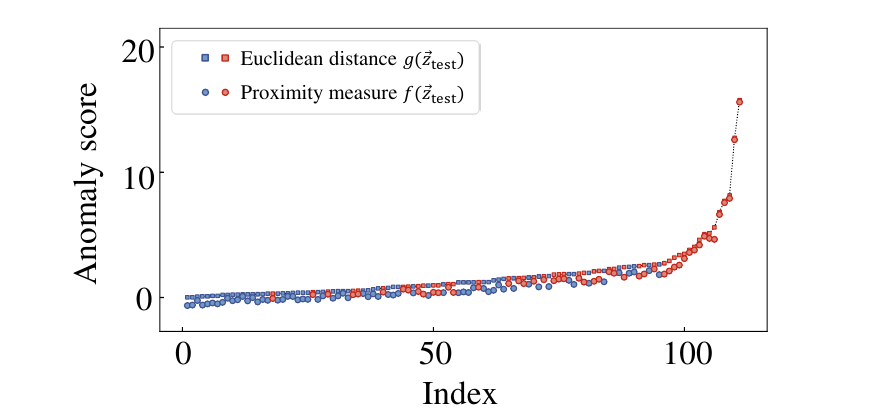}
	\caption{The anomaly scores of all the experimentally measured audio samples in the test set. The audio samples of violin are shown as blue and other types are shown as red.}\label{figs2}
\end{figure}

In the experiments, we train the quantum machine with 4 audio samples of the violin and then perform proximity measurement on \textbf{111} test samples in the dataset, with results shown in Fig.~\ref{figs2}. All the samples are re-ordered in the ascending order of their Euclidean distance $g\left(\vec{z}_{\rm test}\right)$ from the centroid of the data. Both Euclidean distance $g\left(\vec{z}_{\rm test}\right)$ and proximity measure $f\left(\vec{z}_{\rm test}\right)$ of each test sample are shown. 

\end{document}